\documentclass{iau} 
\usepackage{graphicx}

\title[] 
{Deriving the solar activity cycle modulation on cosmic ray intensity observed by Nagoya muon detector from October 1970 until December 2012}

\author[]   
{Rafael R.S. de Mendon\c{c}a$^1$, Carlos. R. Braga$^1$, Ezequiel Echer$^1$, Alisson Dal Lago$^1$, Marlos Rockenbach$^1$, Nelson J. Such$^2$ \and Kazuoki Munakata$^3$}

\affiliation{$^1$Space Geophysics Division, National Institute for Space Research,\\
 S\~ao Jos\'e dos Campos, SP, Brazil, \\ 
 email: {\tt rafael.mendonca@inpe.br} \\[\affilskip]
$^2$Southern Regional Space Research Center (CRS/INPE), \\ 
P.O. Box 5021, 97110-970, Santa Maria, RS, Brazil \\
$^3$Physics Department, Shinshu University, \\ 
Matsumoto, Nagano 390-8621, Japan}

\pubyear{2016}
\volume{328}  
\jname{Living around active stars}
\editors{D. Nandi, A. Valio \& P. Petit, eds.}
\begin{document}

\maketitle

\begin{abstract}
It is well known that the cosmic ray intensity observed at the Earth’s surface presents an 11 and 22-yr variations associated with the solar activity cycle. However, the observation and analysis of this modulation through ground muon detectors datahave been difficult due to the temperature effect. Furthermore, instrumental changes or temporary problems may difficult the analysis of these variations. In this work, we analyze the cosmic ray intensity observed since October 1970 until December 2012 by the Nagoya muon detector. We show the results obtained after analyzing all discontinuities and gaps present in this data and removing changes not related to natural phenomena. We also show the results found using the mass weighted method for eliminate the influence of atmospheric temperature changes on muon intensity observed at ground. As a preliminary result of our analyses, we show the solar cycle modulation in the muon intensity observed for more than 40 years. 
\end{abstract}

\firstsection 
\section{Introduction}

Cosmic rays are charged particles (mostly protons) with energy from MeV to ZeV (10$^{21}$ eV) that travel in space and hit the Earth in an almost isotropic flow. They respond to the configuration of the Interplanetary Magnetic Field (IMF) presenting anisotropies related to solar and interplanetary (transient or recurrent) phenomena (\cite [Potgieter 2013]{Potgieter2013}). Thus, studying the anisotropies in the cosmic ray fluxuseful for understanding the physical aspects of solar and interplanetary phenomena, which can make cosmic rays a useful tool for predicting and monitoring the Space Weather conditions (\cite [Bieber \& Evenson 1998]{BieberEvenson98}; \cite[Munakata et al. 2003]{Munakata_etal03}; \cite [Kudela et al. 2000]{Kudela_etal00}). In this work, we show the procedures performed to making possible to observe clearly the 11 and 22-yrs cosmic ray intensity variations related to the solar activity cycle on the Nagoya muon vertical detector directional channel data recorded between October 1970 and December 2012.
\firstsection
\section{Nagoya muon detector and its data analysis}
\begin{figure}[ht]
\vspace*{-0.4 cm}
\begin{center}
\includegraphics[scale=0.2]{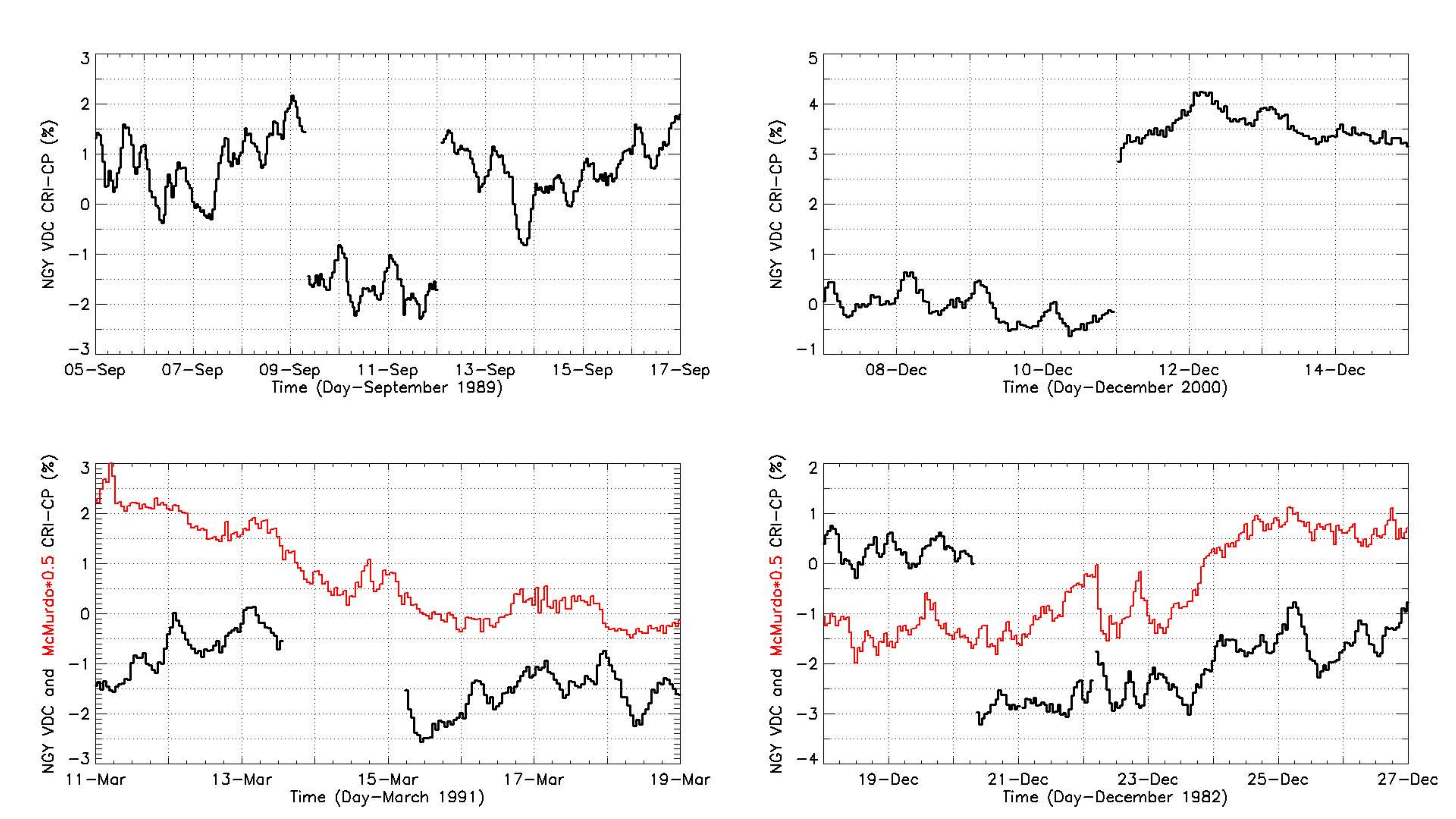} 
\vspace*{-0.4 cm}
\caption{Examples of discontinuities and gaps found on the original hourly Nagoya (NGY) detector Vertical Directional Channel (VDC) Cosmic Ray Intensity Corrected by Pressure effect (CRI-CP).The black curve shows the NGY data, while the red curve represents the McMurdo neutron monitor data multiplied by 0.5.}
\label{fig1}
\end{center}
\end{figure}

The Nagoya (NGY) muon detector is part of the Global Muon Detector Network (GMDN) and is located in Nagoya - Japan (35.15$^{\circ}$ N, 136.97$^{\circ}$ E) at 77 m above sea level. The Vertical Directional Channel (VDC) of this detector has a detection area of 36 m$^2$ and observes secondary cosmic ray (muons) arriving with zenith angle lower than 30$^{\circ}$ since 1970. However, many discontinuities and gaps are present in the NGY VDC data actually. As shown in the two upper panels of Fig. \ref{fig1}, most of discontinuities can be easily identified as caused by a problem or change in the detector electronics because there are no natural phenomena capable of generating variations similar to them. However, it is necessary to take into account that not every change which occurs together a discontinuity or gap is due to that situation. For example, comparing data from Nagoya detector and McMurdo Neutron monitor, we can see that the decrease associated with the gap occurred on March 1991 (black curve in the lower left panel of Fig. \ref{fig1}) is real because a decrease is also present in McMurdo data (red curve). On the other hand, the decrease associated with the gap observed on December 1982 (black curve in the right panel on the bottom part of Fig. \ref{fig1}) is not present on McMurdo data suggesting that is not caused by a natural phenomenon. It is important to stand out that we used neutron monitor, geomagnectic and interplanetary data only for decide if a adjustment on NGY data is necessary or not. The coefficients used for correcting them are calculated taking into account NGY data only. 
\par
\begin{figure} [ht]
\vspace*{-0.4 cm}
\begin{center}
 \includegraphics[scale=0.2]{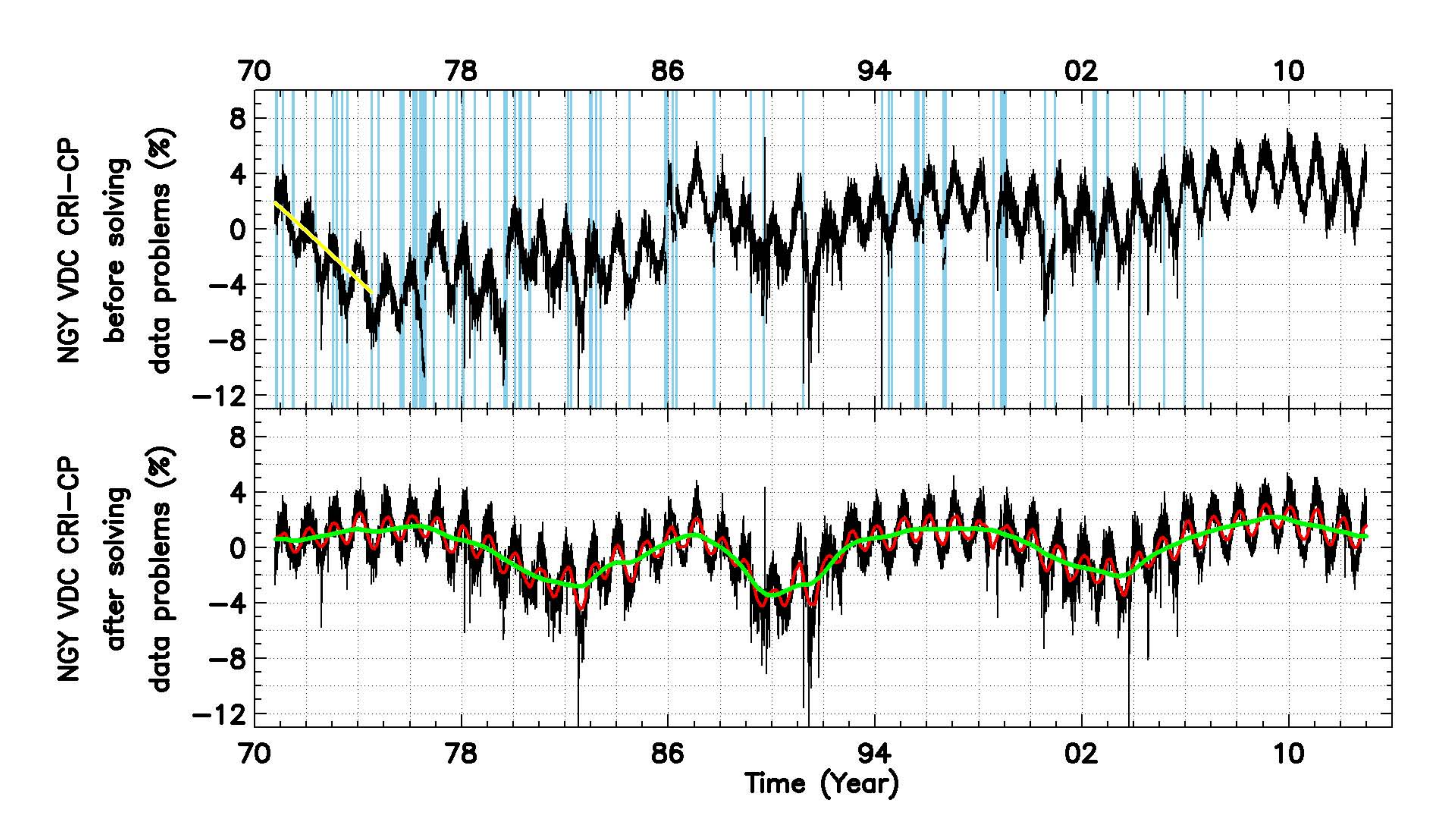} 
 \vspace*{-0.5 cm}
 \caption{Hourly Nagoya Vertical Directional Channel Cosmic Ray IntensityCorrected by Pressure before and after solving data problems (black curves). The vertical light blue lines in the top panel indicate the periods when we notice non-natural changes in CRI-CP. The yellow slanted line in the top panel highlights the period when we assumed a detection efficiency decrease. The red and green curves in bottom panel show the 6 and 12-month running averages of NGY data after eliminating non-natural variations.}
   \label{fig2}
\end{center}
\end{figure}

 The top panel of Fig. \ref{fig2} shows the hourly mean cosmic ray intensity corrected by pressure observed by NGY VDC from October 1970 until December 2012 without any bad data removal or level adjustment. The vertical light blue lines highlight the periods where non-natural changes are found after analyzing this data considering (when available) the cosmic ray intensity variation observed by other detectors together with parameters related to the interplanetary medium and the geomagnetic field (Dst Index and ACE spacecraft data). We believe that the huge intensity decrease occurred between 1970 and 1974 (highlighted by the slanted yellow line) is related to an inital efficiency detection decrease, since there are no known natural phenomena that can produce this monotonous decrease. After we solved these data problems, we can observe two clear periodic variations with periods around one and eleven years (see red and green curves in the bottom panel of Fig. \ref{fig2}).
\par
\begin{figure} [ht]
\vspace*{-0.4 cm}
\begin{center}
 \includegraphics[scale=0.2]{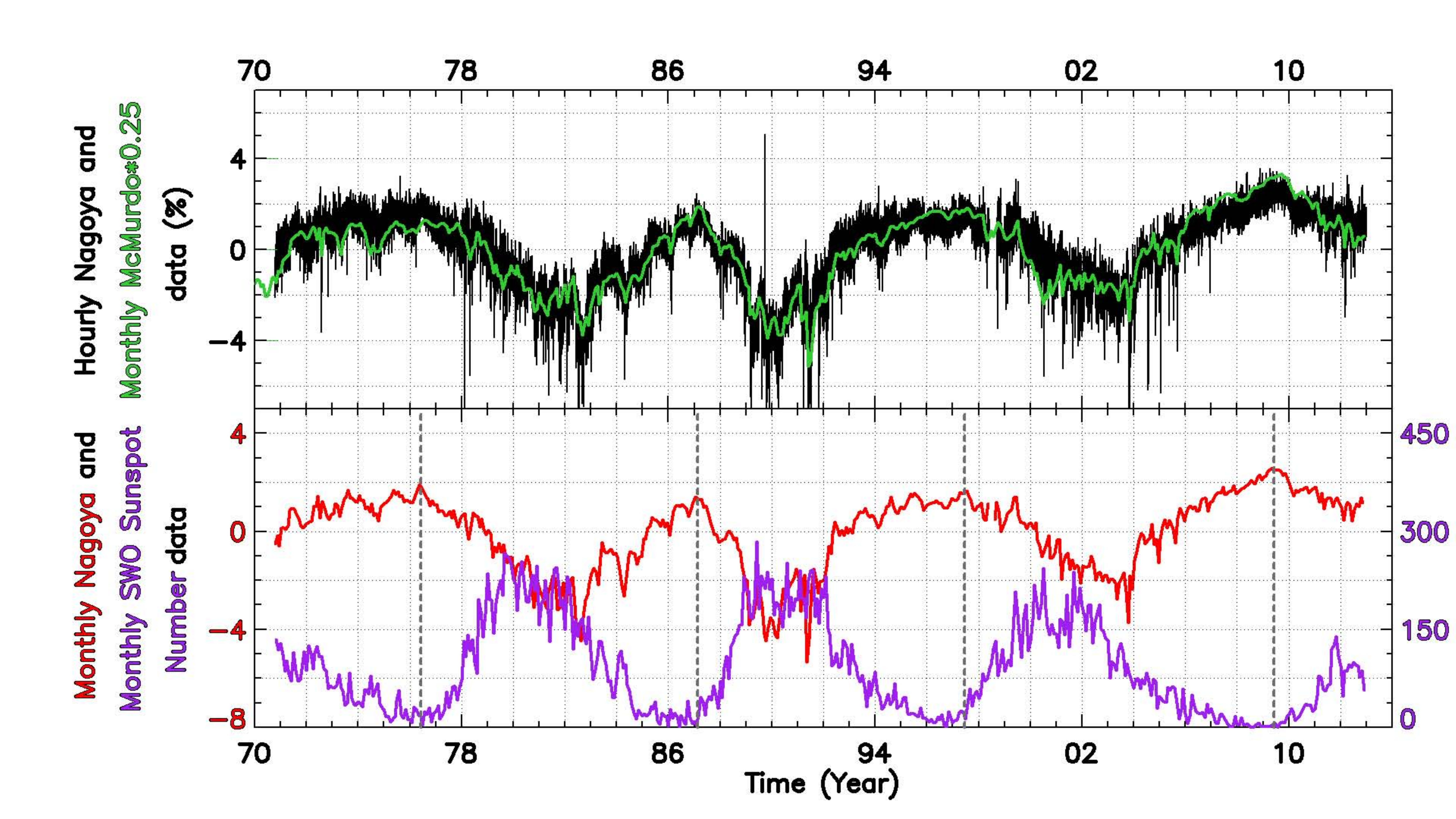} 
 \vspace*{-0.5 cm}
 \caption{Hourly and Monthly NGY VDC cosmic ray intensity after solving electronic data problems and eliminating the temperature effect (black curve). The green curve shows the monthly average McMurdo neutron monitor data multiplied by 0.25. The red curve  shows the monthly mean of NGY data. The purple curve represents the Monthly SWO Sunspot number.}
   \label{fig3}
\end{center}
\end{figure}

 The clear seasonal variation seen in the black and red curves on Fig. \ref{fig2}  is related to the temperature effect that influences the muon creation and disintegration processes in the cosmic ray shower (\cite [Sagisaka 1986]{Sagisaka86}). For ground muon detectors (like NGY), we expect an anti-correlation between the observed cosmic ray count rate and the atmospheric temperature changes. Because of the atmospheric expansion on summer, most of muons are generated at higher altitude having a longer path to cross before reaching the detector at ground. This allows more of them to decay causing a decrease on its intensity at surface (\cite [Sagisaka 1986]{Sagisaka86}; \cite[Mendon\c{c}a et al. 2016]{Mendonca_etal16}). We removed the temperature effect from the vertical directional channel data of Nagoya muon detector utilizing the Mass Weighted Method and using the methodology shown in \cite[Mendon\c{c}a et al. (2016)]{Mendonca_etal16}. This method considers the whole profile of the atmospheric temperature weighted by the atmospheric air density profile. In other words, the temperature measured in an atmospheric layer is weighted by the air mass quantity present in this layer. Using atmospheric temperature data from the Global Data Assimilation System (GDAS), we obtain a mass weighted temperature coefficient equal to -0.255 \%/K , which is similar to that found by \cite[Mendon\c{c}a et al. (2016)]{Mendonca_etal16}. Using this coefficient, we eliminated the temperature effect from NGY VDC data. After that, as we can see in the top panel of Fig. \ref{fig3}, NGY VCD data present a good correlation with the cosmic ray intensity observed by McMurdo neutron monitor and show a clear and full solar activity cycle modulation (11-yr variation with a peak-plateau alternation on the intensity maxima). Comparing monthly mean data of Nagoya and WSO sunspot number (bottom panel of Fig. \ref{fig3}), we also notice that the maximum cosmic ray intensity in each solar minimum occurs just before the sunspot number sharp increase. 

\firstsection
\section{Summary and Final Remarks}
After eliminating non-natural changes present in Nagoya muon detector vertical channel data and correcting it for the temperature effect, a clear 11 and 22 years variations have been clearly observed in a good correlation (R = 0.953) with the cosmic ray intensity observed by the McMurdo Neutron Monitor. This indicates that the instrumental problems, such as data gaps and discontinuities, and the atmospheric temperature effect are successfully removed in the present analysis.. Analyzing data recorded in the last three solar cycles, a clear anti-correlation (R = -0.746) between the solar activity cycle and the cosmic ray intensity observed by Nagoya muon detector corrected was found. These results does not be induced by the data problems correction since it is found in periods with no big problems in the data are found. If we analyze the data (corrected by temperature) between 1986 and 2012 correcting only the huge data problem occurred in 2001, we obtain a satisfactory correlation with NM data and anti-correlation with solar activy for example. Detailed analysis of the anti-correlation with the solar activity will be done in a future work.

\begin{acknowledgements}
The authors acknowledges the LOC of the IAUS 328, the Neutron Monitor Database and the institutions responsible to provide Dst Index and ACE data, CNPq for grants 152050/2016-7, 304209/2014-7, 302583/2015-7 and FAPESP for grant 2014/24711-6. The observations with the Nagoya muon detector are supported by Nagoya University. The McMurdo neutron monitor is supported by National Science Foundation award OPP-0739620. 
\end{acknowledgements}

\end{document}